\documentclass[twocolumn,superscriptaddress]{revtex4}

\usepackage{graphicx}

\begin{document}

\title{On a Type of Self-Avoiding Random Walk with Multiple Site Weightings and Restrictions}

\author{J.\ Krawczyk} \email{j.krawczyk@ms.unimelb.edu.au}
\affiliation{Department of Mathematics and Statistics, The University of Melbourne, 3010, Australia}
\author{T.\ Prellberg} \email{t.prellberg@qmul.ac.uk}
\affiliation{School of Mathematical Sciences, Queen Mary, University of London, Mile End Road, London E1 4NS, United Kingdom}
\author{A.\ L.\ Owczarek} \email{a.owczarek@ms.unimelb.edu.au} 
\affiliation{Department of Mathematics and Statistics, The University of Melbourne, 3010, Australia}
\author{A.\ Rechnitzer} \email{a.rechnitzer@ms.unimelb.edu.au} 
\affiliation{Department of Mathematics and Statistics, The University of Melbourne, 3010, Australia}

\begin{abstract}

We introduce a new class of models for polymer collapse, given by random walks 
on regular lattices which are weighted according to multiple site visits.
A Boltzmann weight $\omega_l$ is assigned to each $(l+1)$-fold visited lattice site, and
self-avoidance is incorporated by restricting to a maximal number $K$ of visits to any site via
setting $\omega_l=0$ for $l\geq K$.
In this paper we study this model on the square and simple cubic lattices for the case $K=3$.
Moreover, we consider a variant of this model, in which we forbid immediate self-reversal of the
random walk. We perform simulations for random walks up to $n=1024$ steps using FlatPERM, a 
flat histogram stochastic growth algorithm.
Unexpectedly, we find evidence that the existence of a collapse transition depends sensitively
on the details of the model.

\end{abstract}
\maketitle

\section{Introduction}

The transition of a flexible macromolecular chain from a random-coil conformation to a globular 
compact form, called coil-globule transition, has been a subject of extensive theoretical and
experimental studies \cite{baysal2003}. 
Generally, polymers in a good solvent are modelled by random walks with short-range repulsion
(excluded volume). Polymers undergoing a coil-globule transition are then modelled by adding an additional 
short-range attraction. The canonical lattice model \cite{orr1946,bennett1998} for this transition is given by interacting self-avoiding
walks (ISAW), in which self-avoiding random walks on a lattice are weighted according to the number of
nearest-neighbour contacts.

From the point of view of continuum models, the drawback of ISAW is that it contains two different kind
of interactions (on-site and nearest-neighbour). In this paper, we introduce a different class of 
lattice models for polymer collapse, which has only on-site interactions. This is in spirit similar
to the Domb-Joyce model \cite{domb1972}, in which a random walk is weighted according to the number
of multiple visits of lattice sites.

It is generally assumed that any reasonable random-walk model with excluded volume and short-range 
attraction should describe the coil-globule transition. Additionally, if the collapsed globule is a liquid-like
bubble, the transition is expected to be second-order \cite{gennes1979a-a}, and if the collapsed globule is frozen the transition 
is expected to be first-order \cite{grassberger1997b}, at least in three dimensions.

However, an investigation of our new class of models reveals that not only the strength of the coil-globule
transition, but its very existence depends sensitively on details of the model.

\section{The class of models and the algorithm}
We consider $n$-step random walks
$\xi=(\vec\xi_0,\vec\xi_1,\ldots,\vec\xi_n)$ on a lattice.
The number of visits to each site $\vec x$ induces a density
$\phi_\xi$ on the lattice sites $\vec x$ via
\begin{equation}
\phi_\xi(\vec x)=\sum_{i=0}^n\delta_{\vec\xi_i,\vec x}\;.
\end{equation}
Interpreting the density $\phi=\phi_\xi$ as a field induced by a
particular random walk configuration $\xi$, we denote the energy of
the field as $E(\phi)$.
In the Domb-Joyce model, the energy functional is given
by
\begin{equation}
\label{eq_dj}
E_{DJ}(\phi)=a\sum_{\vec x}\phi(\vec x)+b\sum_{\vec x}\phi^2(\vec x)\;.
\end{equation}
The first term in this expression is simply related to the length $n$
of the random walk, as
\begin{equation}
\sum_{\vec x}\phi(\vec x)=n+1\;,
\end{equation} 
so that $a$ is related to a chemical potential.  For $b=0$ we have a
pure random walk, while for $b<0$ the model is weakly self-avoiding.
The case $b>0$ leads to an extremely collapsed phase, which is
dominated by configurations occupying few lattice sites with very high
density. Thus, while this model is capable of modelling the swollen
polymer regime, further terms in the energy functional need to be
taken into consideration to model ``realistic'' polymer collapse. 

Generalizing Eq.\ \ref{eq_dj}, we write the energy for a given configuration $\xi$ as
\begin{equation}
E(\xi)=E(\phi_\xi)=\sum_{\vec x}f\left(\phi(\vec x)\right)\;.
\end{equation}
In Eq.\ \ref{eq_dj}, $f(t)$ is simply the quadratic polynomial
$f(t)=at+bt^2$, and any particular choice of $f(t)$ gives an alternative to
the Domb-Joyce model.

Restricting to a maximal number $K$ of visits to any site incorporates self-avoidance. Choosing $K=1$ gives
self-avoiding walks, and for $K>1$ we obtain a model with $K-1$ parameters.
To be precise, we choose $f$ to be given by $f(0)=f(1)=0$,
\begin{equation}
f(2)=\varepsilon_1\;,\quad f(3)=\varepsilon_2\;,\quad\ldots\quad f(K)=\varepsilon_{K-1}
\end{equation}
and $f(t)=\infty$ for $t>K$. Thus, each $l$-fold visited site contributes $\varepsilon_{l-1}$ to the energy
of a configuration.

The canonical partition function is given by
\begin{equation}
\label{eq_par_f}
Z_n(\beta)=\sum_{|\xi|=n+1}e^{-\beta E(\xi)}\;,
\end{equation}
where the sum extends over all random walk configurations with $n$
steps, respectively $n+1$ sites. 
Writing
\begin{equation}
\vec\varepsilon=(\varepsilon_1,\ldots,\varepsilon_{K-1})\quad\mbox{and}\quad\vec{m}=(m_1,\ldots,m_{K-1})
\end{equation} where $m_l$ denotes the
number of sites which are occupied by $l+1$ monomers, 
the energy can be written as
\begin{equation}
E(\vec m)=\sum_{i=1}^{K-1} \varepsilon_im_i=\vec\varepsilon\cdot\vec m\;.
\end{equation}
This enables us to write the partition function Eq.\ \ref{eq_par_f} as
\begin{equation}
Z_n(\beta)=\sum_{\vec m}C_{n,\vec m}e^{-\beta E(\vec m)}=\sum_{\vec m}C_{n,\vec m}e^{\vec\beta\cdot\vec m}\;
\end{equation}
where $C_{n,\vec m}$ denotes the density of states, and $\vec\beta=(\beta_1,\ldots,\beta_{K-1})$ are
generalized temperature parameters, given by $\beta_l=-\beta\varepsilon_l$. In other words, $(l+1)$-fold
visited sites carry a Boltzmann weight $\omega_l=e^{\beta_l}$, with $\omega_0=1$ and $\omega_l=0$
for $l\geq K$.

The density of states is estimated directly by the FlatPERM algorithm (see below for a description). 
Any averaged quantity $Q$ over the set of parameters $\vec m$ for a given length $n$ is calculated by
\begin{equation}
\label{eq_quant}
\left<Q\right>_n(\vec\beta)=
\frac{\sum\limits_{\vec m}Q_{n,\vec m}C_{n,\vec m}e^{\vec\beta\cdot\vec m}}
{\sum\limits_{\vec m}C_{n,\vec m}e^{\vec\beta\cdot\vec m}}\;.
\end{equation}
For our simulations, we restrict to $K=3$, i.e. we only allow two-fold and three-fold visits to any site,
so that we have two free parameters $\beta_1$ and $\beta_2$.

\begin{figure}[t]
  \includegraphics[width=8cm]{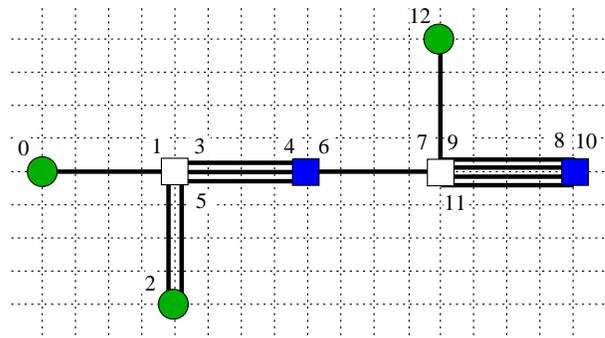}
\caption{\label{fig_configurations} Example of a $12$-step walk on the square lattice with self-reversal allowed (RA).
  A filled circle (green) denotes the presence of a single monomer, filled squares (blue) - two and empty squares (white) 
  - three monomers. The numbers denote the sequence of monomers.}
\end{figure}

We consider two variants of the model which differ in the underlying set of
random walks used. For the first variant, we include all simple random walk configurations,
whereas for the second variant, we only include simple random walks without immediate
self-reversal. For this reason, we call the first variant RA for ``reversal allowed'',
and the second variant RF for ``reversal forbidden''.  Clearly, RF configurations form a subset
of RA configurations. An example of a configuration of the RA model is shown in 
Fig.\ \ref{fig_configurations} for the case of a square lattice. 
We shall consider both models in two dimensions on the square lattice and in three dimensions on
the simple cubic lattice, so that we have a total of four models, which we denote by 
RA2, RA3, RF2, and RF3.

We have simulated these four models using the FlatPERM algorithm \cite{prellberg2004}.  
The power of this algorithm is the ability to sample the density of states uniformly with respect 
to a chosen parametrisation, so that the whole parameter range is accessible from one simulation.

The natural parameters for this problem are $m_1$ and $m_2$. The algorithm directly estimates the
density of states $C_{n,m_1,m_2}$ for all $n\le n_{max}$ and any value of $m_1$ and $m_2$. From this, 
we can then calculate all interesting quantities using Eq.\ \ref{eq_quant}. As we need to store the full 
density of states, we only perform simulations up to a maximal length of $n_{max}=256$.

Fixing one of the parameters $\beta_1$ and $\beta_2$ reduces the size the histogram, and enables us to perform
simulations of larger systems. Fixing $\beta_2$, say, the algorithm directly estimates a partially summed 
density of states
\begin{equation}
\bar{C}_{n,m_1}(\beta_2)=\sum\limits_{m_2}C_{n,m_1,m_2}e^{\beta_2 m_2}\;.
\end{equation}
In this way, we can simulate lengths up to $n_{max}=1024$ at specifically chosen parameters $\beta_1$ or $\beta_2$.
Any averaged quantity $\left<Q\right>_n$ is now calculated by using a suitably modified version of relation \ref{eq_quant}.

\section{Results}
\label{results}

For all four models we find SAW behaviour in the case of repulsion
(i.e. $\beta_1,\beta_2<0$). Here, singly visited sites dominate, and the polymer
is swollen, as is clearly evident from the scaling of the mean-squared end-to-end distance. 

When $\beta_2\ll0\ll\beta_1$, doubly visited sites should dominate, and when 
$\beta_1\ll0\ll\beta_2$, triply visited sites should dominate. Our simulations confirm this, as well.

We now turn to the question of phase transitions between these regimes.
Naively one would expect to find coil-globule transitions from the swollen phase to
the collapsed region. Moreover, for $\beta_1,\beta_2\gg0$, there is competition between
doubly visited and triply visited sites, along with the possibility of a further transition.

We have investigated this scenario in detail for all four models.

\begin{figure}[t]
  \includegraphics[width=8cm]{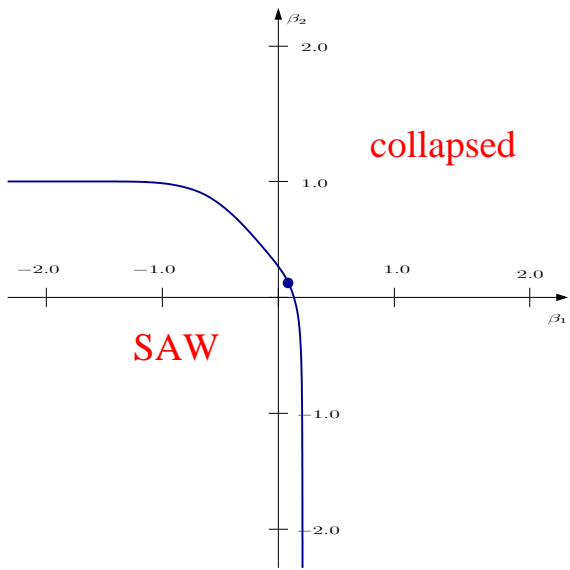} 
\caption{\label{pd_3d_irf}
Model RF3 with two different phase transitions. On varying $\beta_2$ at fixed negative $\beta_1$, there is one type
of transition (possibly first-order), and on varying $\beta_1$ at fixed negative $\beta_2$, there is another. The dot
represents the point at which the type of transition changes.}
\end{figure}

\begin{figure}[t]
 \includegraphics[width=8.5cm]{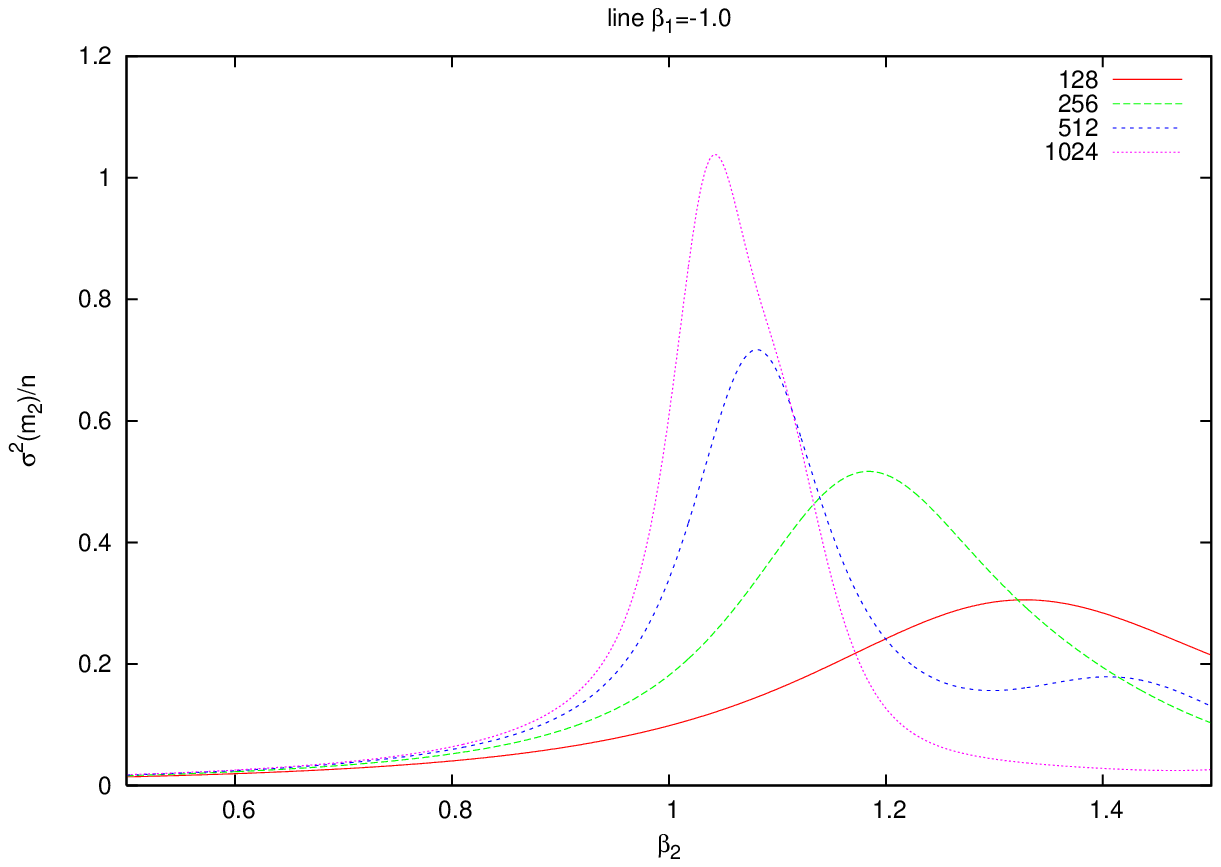}
 \includegraphics[width=8.5cm]{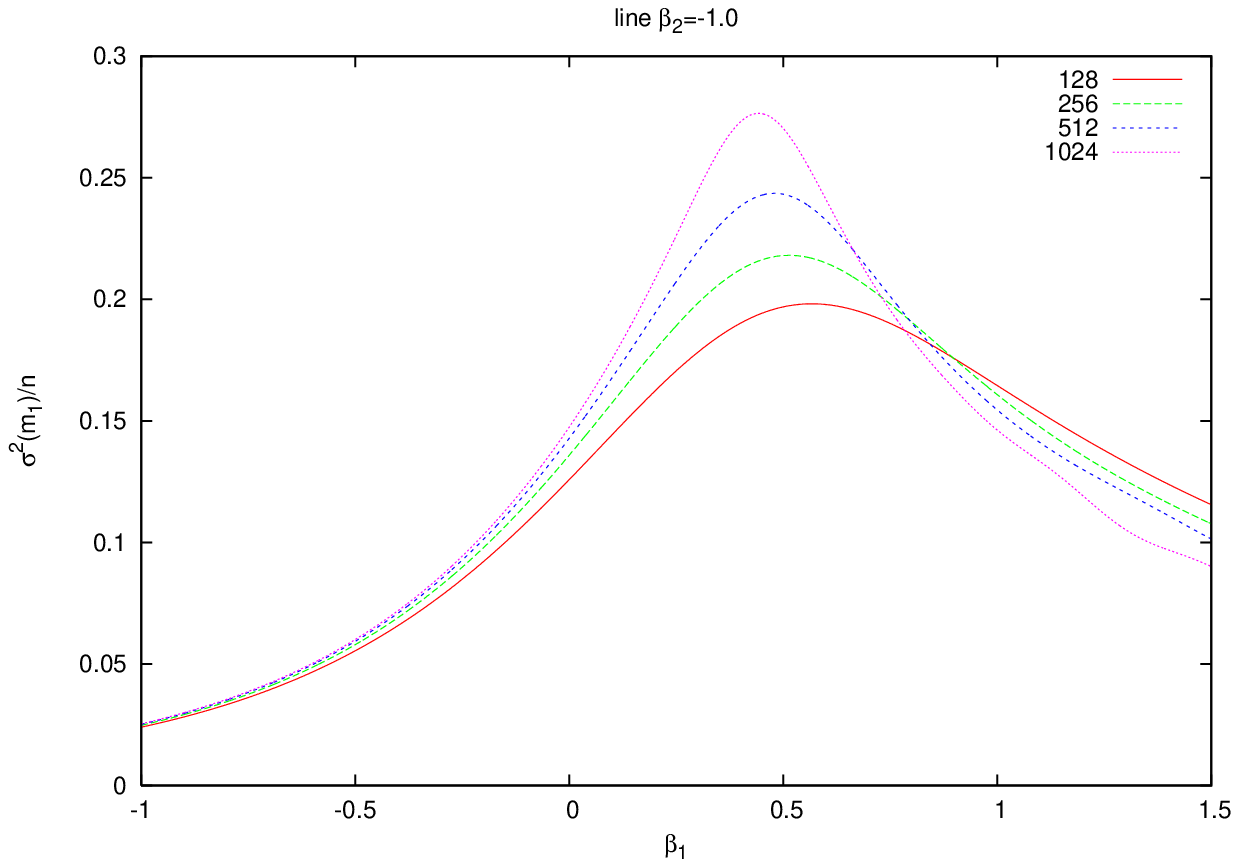} 
\caption{\label{tr_3d_irf} Fluctuations in $m_2$ at $\beta_1=-1.0$ (top) and in $m_1$ at $\beta_2=-1.0$ (bottom) for model RF3.
}
\end{figure}

\begin{figure}[h!]
 \includegraphics[width=8.5cm]{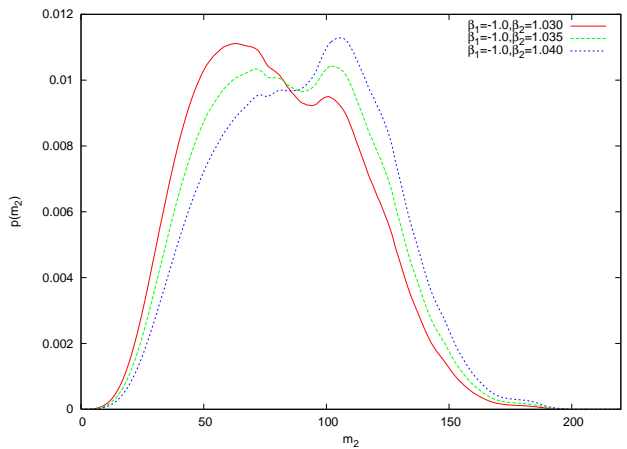}
\caption{\label{bimodal} Distribution of $m_2$ at $\beta_2=-1.0$ near the phase transition for model RF3.
}
\end{figure}

\begin{figure}
  \includegraphics[width=8.5cm]{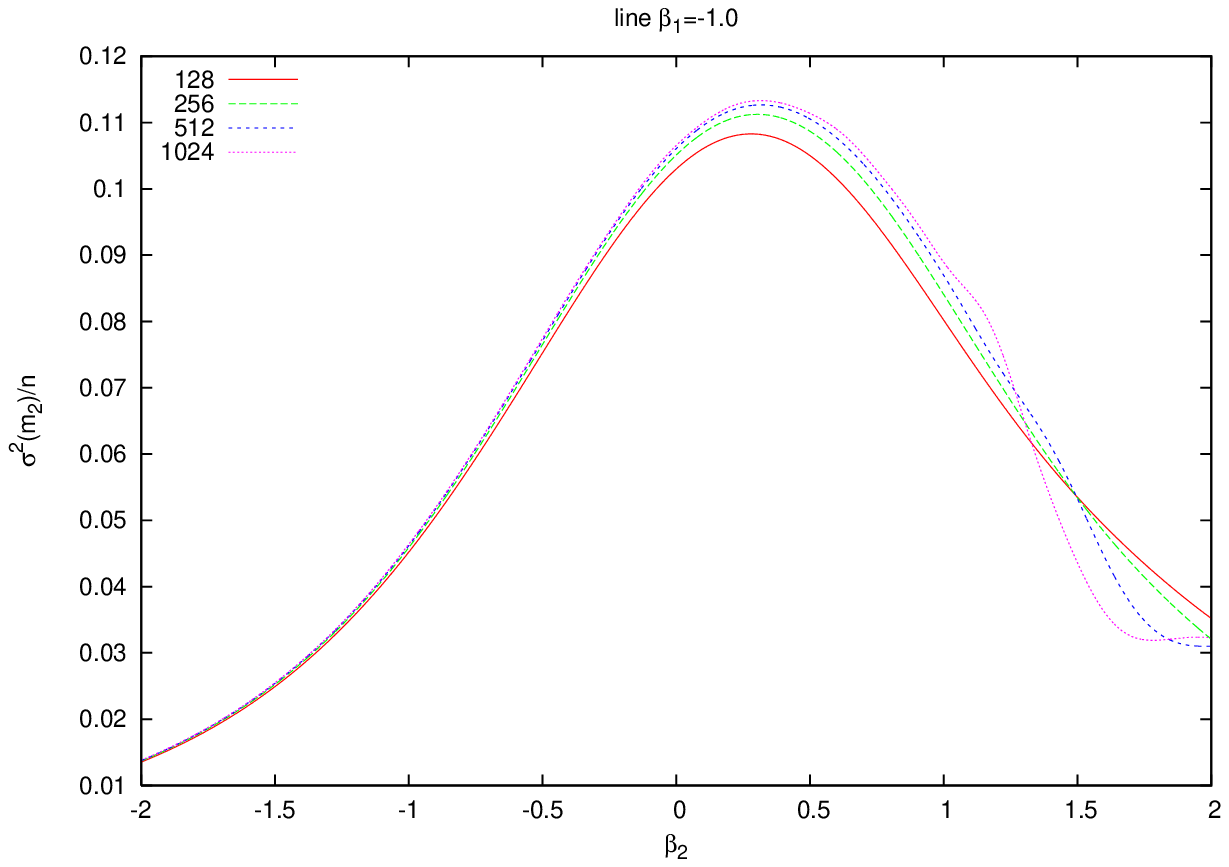}
  \includegraphics[width=8.5cm]{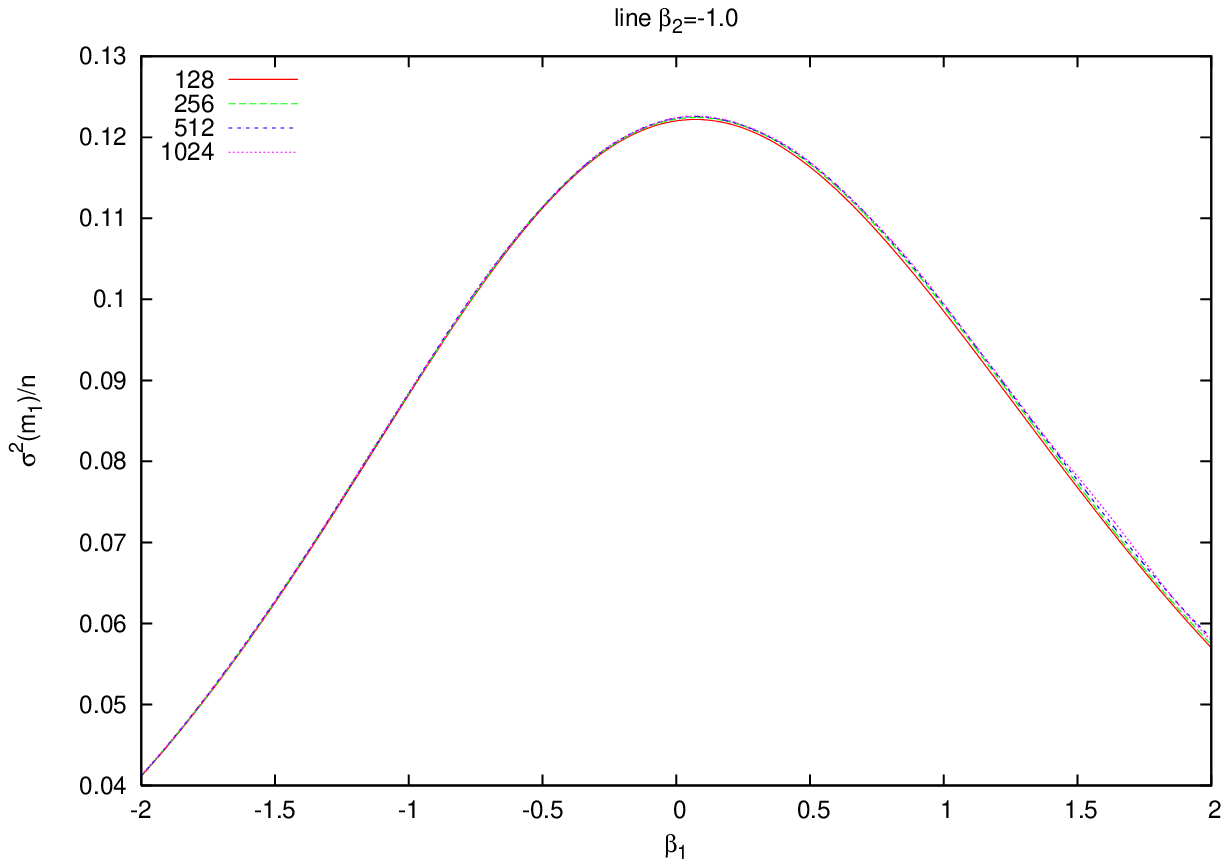} 
\caption{\label{tr_2dira} Fluctuations in $m_2$ at $\beta_1=-1.0$ (top) and in $m_1$ at $\beta_2=-1.0$ (bottom) for model RA2,
showing convergence to smooth thermodynamic functions.
}
\end{figure}


\subsection{RF3}

For random walks with forbidden reversal on the simple cubic lattice (RF3), we find clear evidence of two different phase 
transitions, leading to the phase diagram sketched in Fig.\ \ref{pd_3d_irf}. We cannot precisely locate the point where the 
two phase transition lines meet, however, it is likely that this point is located in the first quadrant.

We have analysed these two phase transitions from simulations at $\beta_1=-1.0$ and $\beta_2=-1.0$, respectively.
Fig.\ \ref{tr_3d_irf} shows fluctuations in $m_1$ along $\beta_2=-1.0$ and fluctuations
in $m_2$ along $\beta_1=-1.0$. In both cases, there is a buildup of fluctuations as the system size increases. The transition at
fixed $\beta_2=-1.0$ is stronger than the transition at fixed $\beta_1=-1.0$. While the latter transition is second-order, the former
appears to be first-order. 
It may be the case that the latter transition is of the same type as ISAW collapse in three dimensions. 
The first-order character of the former transition is supported by the fact that the distribution of $m_2$ near the transition shows a weak bimodality, 
see Fig.\ \ref{bimodal}. An investigation of the scaling behaviour of the mean-squared end-to-end distance supports these conclusions.

There is no indication of any collapse-collapse transition in the first quadrant 
joining up with the point at which the type of the collapse transition changes.


\subsection{RA2}

We now consider random walks with allowed reversal on the square lattice (RA2), since it provides the largest contrast with RF3.
Surprisingly, for RA2, we do not find {\em any} indication of a phase transition,
but merely a smooth crossover. Fig.\ \ref{tr_3d_irf} shows fluctuations in $m_1$ along $\beta_2=-1.0$ and fluctuations
in $m_2$ along $\beta_1=-1.0$. In both cases, there is a smooth crossover, and no buildup of fluctuations as the system size 
increases.  There could, of course, still be a weak transition. However, an investigation of the scaling behaviour of the 
mean-squared end-to-end distance supports the conclusion of no transitions. At the three points
$(\beta_1,\beta_2)=(-1.0,-1.0)$, $(-1.0,1.0)$, and $(1.0,-1.0)$, we find clear evidence for self-avoiding walk scaling behaviour.
We conclude that RA2 is in the self-avoiding walk universality class for all values of $\beta_1$ and $\beta_2$.

So it would seem that changing the dimension and allowing for reversals has removed the phase transition altogether. 
This is unexpected.\\


\subsection{RA3/RF2}

Our analysis of the two remaining models shows that these in some way interpolate between RF3 and RA2.
Random walks with allowed reversal on the simple cubic lattice (RA3) and random walks
with forbidden reversal on the square lattice (RF2) show behaviour similar to each other.

For negative values of $\beta_1$, we find a transition from a swollen to a collapsed phase upon increasing $\beta_2$.
However, for negative values of $\beta_2$, we cannot decide whether there exists a very weak phase transition (the specific heat
exponent $\alpha$ may be negative) or a simple crossover. An analysis of the mean-squared end-to-end distance scaling is inconclusive.

\section{Conclusion}

In conclusion, we have introduced and simulated various new models of polymer collapse in two and three dimensions. We have found 
evidence that the type and very existence of the transition depends crucially on subtle aspects of the underlying lattice model,
in particular on whether the random walk contains immediate reversals or not. There is clearly need for further work to be done to
understand these intriguing results.
If backed up, these results will surely challenge the current theoretical framework of our understanding of polymer collapse.

 Financial support from the Australian Research Council and the Centre
 of Excellence for Mathematics and Statistics of Complex Systems is
 gratefully acknowledged by the authors. They also thank the DFG for financial support.

\end{document}